\documentclass[sigconf, nonacm]{acmart}

\usepackage[table]{xcolor}
\usepackage{algorithmicx}
\usepackage{algpseudocode}
\usepackage{amsmath} 
\usepackage{mdframed}
\usepackage{graphicx}
\usepackage{xcolor}
\usepackage{graphicx}
\usepackage{listings}
\usepackage[capitalize,noabbrev]{cleveref}
\usepackage{enumitem}
\usepackage{graphicx}
\usepackage{makecell} 
\usepackage{array}   
\usepackage{booktabs} 
\usepackage{multirow}
\usepackage{tabularx}

\usepackage[ruled,vlined,linesnumbered]{algorithm2e}
\newcommand{\ourmodel}{{\textsc{VeritasFi}}}
\newcommand{\ourmodelb}{{\textbf{\textsc{VeritasFi}}}}

\newcommand{\lcc}[1]{\tcc{\textbf{#1}}}
\usepackage{tabularx}

\definecolor{BlueGreen}{RGB}{0, 150, 150}
\newcommand{\gains}[1]{\textcolor{BlueGreen}{\tiny{($\uparrow$ #1)}}}   % 增加
\newcommand{\neutral}{\textcolor{gray}{\tiny{(=)}}}   % No change
\newcommand{\gainsdeep}[1]{\textcolor{green!40!black}{\tiny{($\uparrow$ #1)}}} % Stage 2: vs Stage 1
\newcommand{\loss}[1]{\textcolor{black}{\tiny{($\downarrow$ #1)}}} % loss
\AtBeginDocument{%
  }

\setcopyright{none}
\begin{document}

\title{\includegraphics[height=20pt]{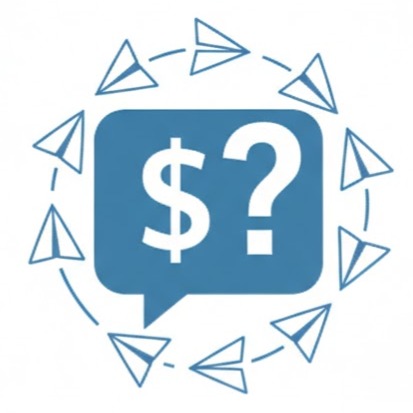}\ourmodel{}: An Adaptable, Multi-tiered RAG Framework for Multi-modal Financial Question Answering}

\author{
Zhenghan Tai$^{1,2 *}$, 
Hanwei Wu$^{1,3 *}$, 
Qingchen Hu$^{1,4 *}$, 
Jijun Chi\textsuperscript{1,2 \textdagger},
Hailin He \textsuperscript{1 \textdagger},
Lei Ding \textsuperscript{1,5 \textdagger}, \\
Tung Sum Thomas Kwok$^{1,6}$,
Bohuai Xiao$^{1}$,
Yuchen Hua$^{1}$,
Suyuchen Wang$^{7,8}$,
Peng Lu$^{7}$,
Muzhi Li$^{9}$,
Yihong Wu$^{7}$,
Liheng Ma$^{4,8}$,
Jerry Huang$^{7,8}$,
Jiayi Zhang$^{10}$,
Gonghao Zhang$^{11}$,
Chaolong Jiang$^{1}$, \\
Jingrui Tian$^{6}$,
Sicheng Lyu$^{1,4,8}$,
Zeyu Li$^{1,12}$,
Boyu Han$^{1,13}$,
Fengran Mo$^{7}$,
Xinyue Yu$^{7,8}$,
Yufei Cui$^{4,8}$,
Ling Zhou$^{14}$,
Xinyu Wang \textsuperscript{1,4 \S},
}

\affiliation{
    \institution{
    $^{1}$SimpleWay.AI \quad
    $^{2}$University of Toronto \quad
    $^{3}$McMaster University \quad
    $^{4}$McGill University \quad
    $^{5}$University of Manitoba \quad
    $^{6}$University of California, Los Angeles\quad
    $^{7}$University of Montreal \quad
    $^{8}$Mila \quad
    $^{9}$CUHK\quad
    $^{10}$HKUST(GZ) \quad
    $^{11}$flab.ai \quad
    $^{12}$Nanyang Technological University \quad
    $^{13}$Stanford University \quad
    $^{14}$CG Matrix Technology Limited
    }
    \country{}
}

\thanks{\textsuperscript{*}Equal Contribution. - \\winfred.tai@mail.utoronto.ca,wuh92@mcmaster.ca, qingchen.hu@mail.mcgill.ca
        \par\textsuperscript{\textdagger}Core Contribution.
        \par\textsuperscript{\S} Corresponding Author - xinyu.wang5@mail.mcgill.ca
        }
% \\
% $^{1}$SimpleWay.AI \quad
% $^{2}$University of Toronto \quad
% $^{3}$McMaster University \quad
% $^{4}$McGill University \quad
% $^{5}$University of Manitoba \quad
% $^{6}$University of California, Los Angeles\quad
% $^{7}$University of Montreal \quad
% $^{8}$Mila \quad
% $^{9}$CUHK\quad
% $^{10}$HKUST(GZ) \quad
% $^{11}$flab.ai \quad
% $^{12}$Nanyang Technological University \quad
% $^{13}$Stanford University \quad
% $^{14}$CG Matrix Technology Limited
% }
% \thanks{*These authors contributed equally to this research.}

% % \author{G.K.M. Tobin}
% \email{xinyu.wang5@mail.mcgill.ca}
% \email{winfred.tai@mail.utoronto.ca}
% \affiliation{%

\renewcommand{\shortauthors}{Zhenghan Tai et al.}

\begin{abstract}
Retrieval-Augmented Generation (RAG) is becoming increasingly essential for Question Answering (QA) in the financial sector, where accurate and contextually grounded insights from complex public disclosures are crucial.
However, existing financial RAG systems face two significant challenges: (1) they struggle to process heterogeneous data formats, such as text, tables, and figures; and (2) they encounter difficulties in balancing general-domain applicability with company-specific adaptation.
To overcome these challenges, we present \textbf{\ourmodel{}}, an innovative hybrid RAG framework that incorporates a multi-modal preprocessing pipeline alongside a cutting-edge two-stage training strategy for its re-ranking component.
\ourmodel{} enhances financial QA through three key innovations: (1) \textbf{A multi-modal preprocessing pipeline} that seamlessly transforms heterogeneous data into a coherent, machine-readable format. (2) \textbf{A tripartite hybrid retrieval engine} that operates in parallel, combining deep multi-path retrieval over a semantically indexed document corpus, real-time data acquisition through tool utilization, and an expert-curated memory bank for high-frequency questions, ensuring comprehensive scope, accuracy, and efficiency. (3) \textbf{A two-stage training strategy} for the document re-ranker, which initially constructs a general, domain-specific model using anonymized data, followed by rapid fine-tuning on company-specific data for targeted applications.
By integrating our proposed designs, \ourmodel{} presents \textbf{a groundbreaking framework} that greatly enhances the adaptability and robustness of financial RAG systems, providing a scalable solution for both general-domain and company-specific QA tasks. Code accompanying this work is available at https://github.com/simplew4y/VeritasFi.git.

\end{abstract}





\maketitle

\section{Introduction}

The integration of Large Language Models (LLMs) into the financial sector has catalyzed the development of sophisticated Question Answering (QA) systems, with Retrieval-Augmented Generation (RAG) rapidly emerging as a cornerstone technology \citep{lewis2021retrievalaugmentedgenerationknowledgeintensivenlp, gao2024retrievalaugmentedgenerationlargelanguage}.

By augmenting LLMs with external knowledge from complex corporate disclosures, such as U.S. Securities and Exchange Commission (SEC) filings, RAG frameworks aim to deliver accurate, verifiable, and context-aware insights essential for high-stakes decision-making and regulatory compliance\citep{chen2022finqadatasetnumericalreasoning, lai2024sec, wang2025fintagging, daimi2024scalable, guo2024lightrag}.
For example, when financial analysts query a company's 10-K report for "risks associated with supply chain disruptions in the last fiscal year", they need more than a simple keyword match. They require a synthesized response that integrates complex textual descriptions, numerical data from tables, and potentially trends depicted in figures.

However, deploying general-purpose RAG in finance remains difficult. Two gaps persist: \textbf{(i)} handling the \emph{multi-modal} nature of filings (text/tables/figures), and \textbf{(ii)} balancing general-domain knowledge with entity-specific adaptation. Many RAG pipelines and financial extractors linearize documents as flat text, leading to incomplete grounding \citep{gao2024retrievalaugmentedgenerationlargelanguage, loukas2025edgar, 10.1145/2797115.2797120, chatterjee2024dreqdocumentrerankingusing}. Domain shift is well documented: zero-shot general rerankers degrade across datasets \citep{thakur2021beir}, while in-domain query generation and finetuning improve reranking \citep{dai2022promptagator, bonifacio2022inpars}; vertical models further reinforce this (biomedical \citep{lee2020biobert, gu2021pubmedbert}, legal \citep{chalkidis2020legalbert, chalkidis2022lexglue}, finance \citep{araci2019finbert}).

To overcome these fundamental limitations, we introduce \ourmodel{}, an adaptable, hybrid RAG framework designed for robust and precise multi-modal financial question answering. At its core, \ourmodel{} combines a multi-modal preprocessing pipeline with a Tripartite Hybrid Retrieval (THR) engine and a two-stage Domain-to-Entity re-ranking strategy. In extensive experiments, \ourmodel{} attains the best end-to-end performance across diverse financial QA datasets, outperforming strong baselines—including GraphRAG and LightRAG—and remaining superior even when classical retrievers (BM25/Faiss) are strengthened with our CAKC preprocessing. Beyond end-to-end accuracy, our multi-path retrieval delivers higher evidence hit rates, while the two-stage re-ranking reliably improves retrieval quality and enables rapid, low-overhead adaptation to new companies.

In summary, our key contributions are as follows:
\begin{itemize}[leftmargin=*]
    \item We propose \ourmodelb{}, a novel, end-to-end RAG framework that integrates multi-modal data processing, hybrid retrieval, and an adaptable re-ranker to address the challenges of financial QA.
    \item We introduce \textbf{a multi-modal preprocessing pipeline} that effectively unifies disparate data formats from financial filings, which include text, tables, and figures, into a coherent, machine-readable format optimized for retrieval.
    \item We design \textbf{a tripartite hybrid retrieval engine} that combines deep multi-path retrieval, a high-frequency question-matching cache, and external tool integration to ensure comprehensive scope, low latency, and factual timeliness.
    \item We develop \textbf{a two-staged, domain-to-entity adaptation strategy} that first builds a generalized financial re-ranker and then rapidly specializes it for a target entity using automated data annotation, which promises a scalable and rapid solution to model customization for onboarding.
    \item The extensive experiments on both public benchmarks and in-house company dataset showcase that \ourmodel{} significantly outperforms existing RAG architectures in correctness, context relevance, and overall answer quality.
\end{itemize}

% % \renewcommand{\paragraph}[1]{\paragraph{\textbf{#1}}}

\section{System Overview}
Tackling the complexities of domain-specific financial question answering requires a system that can synthesize knowledge from highly heterogeneous sources, including multi-modal documents, structured time-series data, and real-time external information. To this end, we propose an advanced hybrid RAG framework designed to meet these demands. The core components of our system, which operate synergistically, are: \textbf{i) a Context-Aware Knowledge Curation (CAKC)} module to process unstructured financial filings;\textbf{ ii) a Tripartite Hybrid Retrieval (THR)} engine that queries diverse information sources in parallel.
%\textbf{iii) a Domain-specialized Re-Ranking (DR)} module to ensure the retrieved reference is both accurate and relevant. 
% An overview of this integrated architecture is shown in Figure~\ref {fig:complete_ppl}

% \begin{figure*}[h] 
%     \centering  
%     \includegraphics[width=0.9\linewidth]{images/FinSage.png}
%     \caption{The complete \ourmodel{} RAG pipeline.}
%     \label{fig:overall_architecture}
% \end{figure*}

\section{Context-Aware Knowledge Curation} \label{sub: ffp}
\begin{figure*}[h] 
    \centering  
    \includegraphics[width=0.9\linewidth]{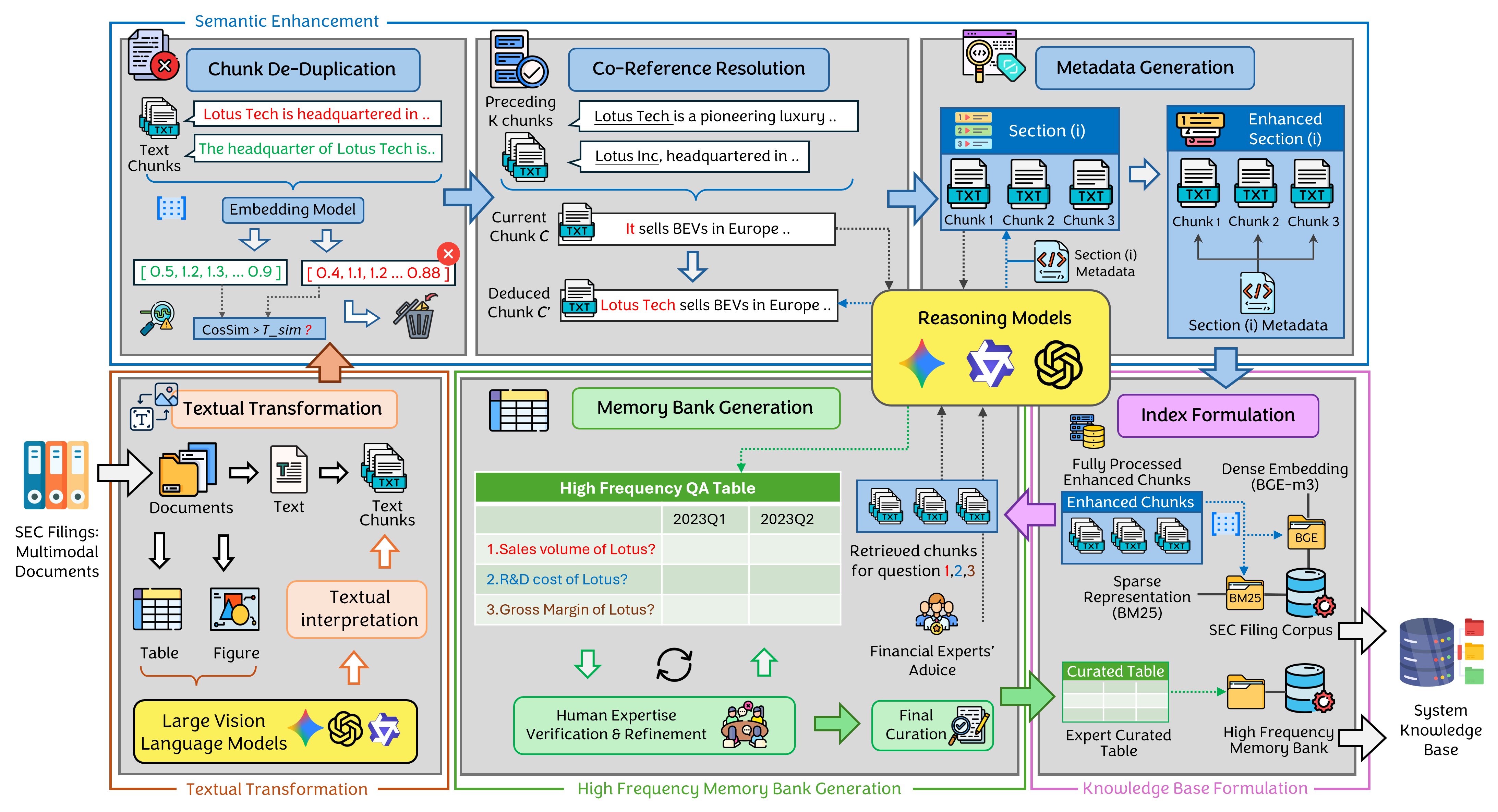}
    \caption{Context-Aware Knowledge Curation (CAKC) pipeline. Textual Transformation Module (Orange) transforms multimodal data into textual representation. Semantic Enhancement module (Blue) refines the text chunks. The Knowledge Base Formulation Module (Purple) generates the final SEC Filing Corpus from processed text chunks, and the High Frequency Memory Bank Generation Module (Green) constructs a High-Frequency Memory Bank of time-stamped answers.}
    \label{fig:FFP}
\end{figure*}

The CAKC module converts raw, multi-modal English financial documents into structured knowledge bases optimized for efficient information retrieval. As shown in Figure \ref{fig:FFP}, its primary goal is to decompose complex filings into discrete, contextually coherent information units that enable accurate and efficient indexing. The pipeline yields two complementary outputs.
\begin{enumerate}[leftmargin=*]
    % \item \textbf{SEC filing Corpus:} Sparse and Dense representations of a corpus of semantically enriched, self-contained text chunks derived from the textual and graphical content in the SEC filings documents.
    \item \textbf{SEC Filing Corpus:} A primary corpus consisting of semantically enriched, self-contained English \textit{text chunks}, transformed and enhanced from SEC filings.
    \item \textbf{High-frequency memory bank:} A structured, factual cache containing \textit{time-stamped answers} to common quantitative questions, derived from a semi-automated, human-in-the-loop process. % It is to ensure high precision for frequently queried data points, serving as a fast-path lookup mechanism.
\end{enumerate}

% This process begins with two key steps for handling unstructured content: \emph{textual transformation} and \emph{semantic enhancement}.

\subsection{Textual Transformation of Multi-Modal Documents}
Our pipeline first parses each multi-modal SEC filing document into a sequence of fixed-length chunks (200 words) with different modality (i.e., text, figure, or table), via the open-source tool MinerU \citep{wang2024mineruopensourcesolutionprecise,he2024opendatalab,wang2024unimernet,zhao2024doclayoutyoloenhancingdocumentlayout}. To create a unified textual representation suitable for LLM, we then employ \texttt{GPT-4o} ~\citep{OpenAI_GPT4o_2024} to transform all non-textual content into descriptive narratives. Specifically, figures are converted into structured captions summarizing their key insights, while tables are rewritten as textual statements that capture their primary data trends and relationships.
\subsubsection{Semantic Enhancements.}
To ensure that the resulting textual chunks are coherent and optimized for retrieval, we perform three semantic enhancements:

\begin{enumerate}[leftmargin=*]
\item \textbf{Similar chunk de-duplication.}
 To reduce content redundancy and improve retrieval diversity,  we conduct pair-wise comparison that discard any chunk whose dense embedding similarity with other chunks exceeds a predefined threshold $\tau_{\text{sim}}$. Specifically, we map each text chunk $c_{i}$ to the dense vector representation of the embedding model $e_{i} = E_{bge\_m3}(c_{i})$, and measure pairwise semantic relationship between text chunks with cosine similarity $\operatorname{CosSim}(\cdot, \cdot)$. The final de-duplicated corpus $\mathcal{C}'$ is then constructed by iteratively removing the subsequent similar chunk $\mathbf{c}_j$:
 % \begin{equation}
 % \mathcal{C}' = \mathcal{C} \setminus \{ c_j \in \mathcal{C} \mid \forall  i \neq j: \text{CosSim}(E_{bge\_m3}(c_i), E_{bge\_m3}(c_j)) > \tau_{\text{sim}} \}
 % \end{equation} 
 \begin{equation}
  \mathcal{C}' = \mathcal{C} \setminus \{ c_j \in \mathcal{C} \mid \forall  i \neq j:  \text{CosSim}(e_i, e_j)) > \tau_{\text{sim}} \},
\end{equation}
This mechanism alleviates redundancy in retrieval by filtering near-identical chunks, effectively conserving the generator’s limited context window.
\item \textbf{Co-reference resolution.}
To enhance the contextual sufficiency of each chunk, we replace the pronouns in a given chunk with their explicit antecedents, as chunks retrieved in isolation lack the necessary context to resolve pronounces, where such ambiguity leads to factual inaccuracies in generation.
\item \textbf{Metadata generation.} 
 To ensure individual chunks retain the broader context from other chunks in
 the same parent section for retrieval, we generate a section-level summary with \texttt{GPT-4o} as a metadata attachment to every chunk. This create a shared contextual anchor that enhances retrieval accuracy for high-level queries. 
\end{enumerate}

\subsection{High Frequency Memory Bank Generation} \label{subsub: high-frequency-memory-bank} 
 Beyond processing unstructured financial filings, the CAKC pipeline pre-caches a High-Frequency Memory Bank to accelerate common quantitative query lookups. We structure the cache as a table where each row corresponds to different canonical, normalized questions curated by financial experts. The columns represent time points (e.g., fiscal quarters), and the cells record the value of question's subject at the specified time point. We initialize the memory bank by running our full RAG pipeline on source filings for each question, where we utilize \texttt{DeepSeek R1} for questions requiring complex reasoning across multiple retrieved chunks. As the initialization is an offline process, we prioritize accuracy by increasing the number of candidate chunks from re-ranking results and employing a reasoning model without latency constraints. The initialized memory bank is then carefully reviewed and verified by in-house experts, resulting in a highly reliable, structured knowledge base for quick answer retrieval. More details regarding the canonical questions and table structure can be found in Appendix~\ref{app:freq_table}.

% $\tau_{\text{bm25}}$
\subsection{Knowledge Base Indexing} 
We index all processed text chunks in parallel for semantic retrieval. Dense representations $e_{i}$ are mapped through $E_{bge\_m3}(\cdot)$ and stored in \texttt{Chroma} vector database \citep{chroma_db}. Sparse representations are build by inverted index over the text chunks. The High-Frequency Memory Bank is indexed in a relational database to allow on-the-fly subsequence and BM25 searches against its canonical questions. The canonical questions are embedded and indexed within the vector datastore to facilitate semantic matching.
\vspace{-0.3cm}

\section{Tripartite Hybrid Retrieval} \label{sub:tripartite-hybrid-retrieval}
\begin{figure*}[h] 
    \centering  
    \includegraphics[width = 0.85\linewidth]{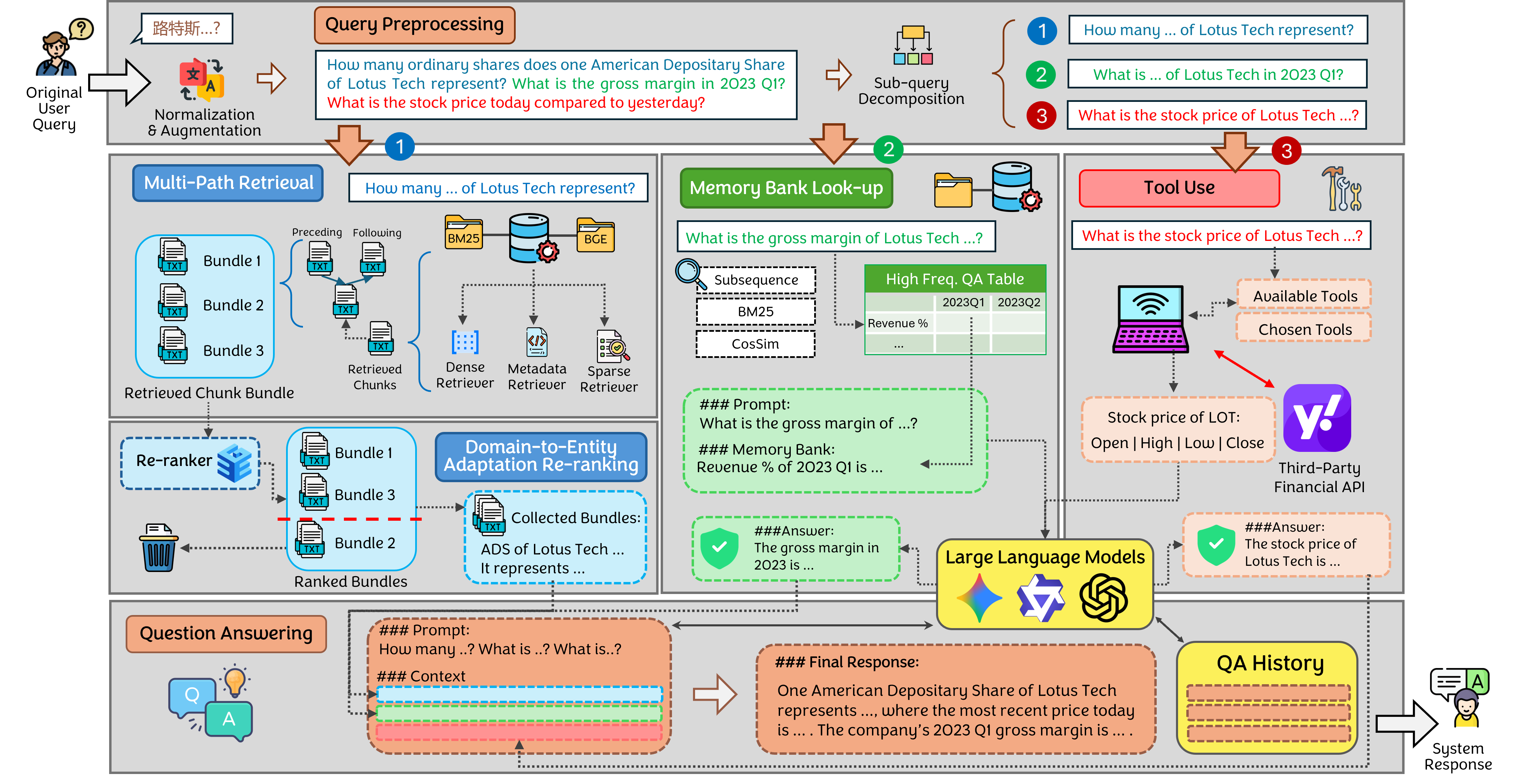}
    \caption{The complete \ourmodel{} RAG pipeline. The diagram shows three concurrent knowledge retrieval paths: Multi-Path Retrieval (Blue), Memory Bank Look-up (Green) and Tool Use (Red).}
    \label{fig:complete_ppl}
\end{figure*}

To provide comprehensive answers, we first run a \emph{Query Preprocessing} stage that normalizes, disambiguates, and decomposes the user query into minimal self-contained sub-queries. Building on this, our \textbf{Tripartite Hybrid Retrieval (THR)} engine (Figure~\ref{fig:complete_ppl}) routes the processed request to three specialized modules in parallel to handle financial queries of different complexities and data types: \textbf{(1) Multi-Path Retrieval with Domain-to-Entity Adaptation Re-ranking} for deep semantic search over unstructured filings; \textbf{(2) High-Frequency Memory Look-up} for rapid answers from a pre-cached memory bank; and \textbf{(3) Tool Use} for real-time data via external APIs. Evidence from all three streams is then fused to synthesize the final answer. 

\subsection{Query Preprocessing}
This module serves as the entry point to the retrieval pipeline, aims to transform complex, multi-faceted financial queries, which rely on unspoken context from previous turns, into precise, machine-actionable instructions. The contextual transformation is achieved in two steps: \begin{enumerate}[leftmargin=*]
\item \textbf{Query Normalization and Context Augmentation} tackles the problem of incompleteness and ambiguity. We normalize the raw query by translating queries into English to align its representation with the semantic space of our English-language corpus. Next, this pipeline includes the entire conversation history for co-reference resolution and context augmentation, allowing LLM to integrate prior turns and identify users' complete intent, especially when current turn is highly elliptical or dependent on previous information. 
\item \textbf{Intent-Driven Decomposition and Retrieval Router} tackles the problem of data complexity and execution efficiency. We decompose enhanced, context-rich query into smallest possible self-contained query units (sub-queries), and assign each to optimal retrieval paths using the High-Frequency Memory Bank, pre-defined Tools, deep retreival from document corpus, or direct answering using LLM's own knowledge. 
\end{enumerate}

\subsection{Multi-Path Retrieval} \label{sub:multi-path-retrieval}
For sub-queries that require in-depth analysis of unstructured financial filings, the system activates the Multi-Path Retrieval and Domain-to-Entity Adaptation Re-ranking stream. This pipeline is designed to maximize recall by simultaneously using diverse retrieval strategies to gather a broad set of candidate chunks.
% Following this initial, a domain-specialized re-ranking module filters and prioritizes these candidates to ensure only the most relevant context is used for generating the final answer.
\subsubsection{Retrieval Module}
The Retrieval Module executes three distinct retrieval strategies in parallel to create a comprehensive candidate set of chunks from the document corpus. Each retriever targets a different aspect of the query-chunk relationship to maximize recall: 
(1) \textbf{BM25 Sparse Retriever:} A lexical search model that ranks chunks using keyword overlap with the query using the Best-Matching 25 (BM25) scoring function.
(2) \textbf{Dense Retriever:} A semantic retriever that represents both queries and document chunks as dense embedded vectors, obtained from an embedding model. Retrieval is performed by computing the cosine similarity between these embeddings, accelerated using FAISS \citep{johnson2017billionscalesimilaritysearchgpus} nearest-neighbor search.
(3) \textbf{Metadata Retriever:} A high-level semantic search that matches the query against LLM-generated summaries attached as metadata to each chunk during the CAKC stage.

\subsubsection{Chunk Bundling Module} 
To capture complete semantic context, this module expands each retrieved chunk $c_i$ by iteratively appending adjacent neighbors based on semantic similarity. For each retrieved chunk, the module starts from an initial bundle $B_i = \{c_i\}$, it examines up to k chunks in both directions. A candidate chunk $c_i\in \{c_{i-k}, \ldots, c_{i-1}, c_{i+1}, \ldots, c_{i+k}\}$ is added to the bundle if 
\begin{equation}
\text{CosSim}(e_i, e_j) > \tau_{\text{bundle}} \quad \text{and} \quad c_j \notin B_i,
\end{equation}
where $e_i= E_{bge\_m3}(c_i)$, and $\text{CosSim}(\cdot, \cdot)$ denotes the cosine similarity function, resulting in the expanded bundle $B_i = B_i \cup \{c_j\}$. This expansion ensures that multi-paragraph discussions in financial documents are captured cohesively.

% To reconstruct the complete semantic context around a search result, this module expands the retrieved chunk with its adjacent neighbors. Since key discussions in financial documents often span multiple paragraphs, the module programmatically evaluates the chunks immediately preceding and following a retrieved chunk, appending them to form a "bundle" if their semantic similarity exceeds a predefined threshold $\tau_{\text{bundle}}$.
% Formally, given a retrieved chunk $c_i$ and a bundle $B_i = \{c_i\}$, the module iteratively examines up to $k$ chunks in both directions. For each candidate chunk $c_j$ in the neighborhood $\{c_{i-k}, \ldots, c_{i-1}, c_{i+1}, \ldots, c_{i+k}\}$, we add $c_j$ to the bundle if:

\subsection{Domain-to-Entity Adaptation Re-ranking}
% To refine the candidate set from the retrieval and bundling stages, the Two-Stage Fine-Tuned Re-ranking (TFR) module acts as the filter to enhance precision. We employ an decoder-only model to score each chunk's relevance to the query, re-ordering the candidates to prioritize the most pertinent information. Unlike traditional cross-encoders that directly output a relevance score, the LLM-based re-ranker operates through a prompt-based binary classification approach \citep{li2023makinglargelanguagemodels}. Given a query $q$ and a candidate chunk $c$, the model receives a concatenated triplet consisting of a structured prompt, the query, and the chunk, and is constrained to produce only binary responses ("yes" or "no") indicating whether the chunk is relevant to the query. Specifically, the re-ranking score is computed as:
% \begin{equation}
% \text{score}(q, c) = \text{softmax}\big(\text{logit}_{\text{"yes"}}, \text{logit}_{\text{"no"}}\big)_{\text{"yes"}},
% \end{equation}
% where $\text{logit}_{\text{"yes"}}$ and $\text{logit}_{\text{"no"}}$ are the logits corresponding to the "yes" and "no" tokens from the model's vocabulary, and the softmax normalization yields the probability of relevance, which serves as the final re-ranking score.

To refine candidates from the retrieval and bundling stages, the Domain-to-Entity Adaptation Re-ranking (DAR) module filters and re-orders chunks to improve precision. A decoder-only model scores each chunk’s relevance to the query using a prompt-based binary classification approach \citep{li2023makinglargelanguagemodels}, rather than producing a direct relevance score like a traditional cross-encoder. Given a query $q$ and candidate chunk $c$, the model classifies the chunk as relevant ("yes") or not ("no"). At inference, the final relevance score is the probability assigned to the "yes" class, computed as:
% \begin{equation}
% \textbf{Score}(q,c) = \text{P}_\text{relevance}(q,c) =\text{softmax}\big(\text{logit}_{\text{[yes]}}, 
% \text{logit}_{\text{[no]}}\big)_{\text{"yes"}}
% \end{equation}
\begin{equation}
\mathrm{Score}(q,c)\;=\; \mathbb{P}(\text{relevant}\mid q,c)
\;=\; \sigma\!\big(z_{\text{yes}} - z_{\text{no}}\big)
%\;=\; \frac{1}{1+\exp\!\big(-(z_{\text{yes}}-z_{\text{no}})\big)}.
\label{eq:score},
\end{equation}

where $z_{\text{yes}}$ and $z_{\text{no}}$ are the logits for “yes”/“no,” and $\sigma(\cdot)$ is the sigmoid.
%During training, we directly use the logit of the "yes" and "no" tokens as the input to our loss function where %The TFR module is optimized with a unified objective that drives the model to distinguish relevant from irrelevant information. 
% we adopt a supervised contrastive learning formulation, enabling the re-ranker to assign higher scores to truly relevant chunks and strengthen its discrimination across general and entity-specific domains.
% where $\text{logit}_{\text{"yes"}}$ and $\text{logit}_{\text{"no"}}$ are the logits for the respective tokens in the model's vocabulary, and the softmax normalization yields the probability of relevance, which serves as the final re-ranking score.
% During training, the model is optimized using a supervised contrastive learning objective computed directly from the "yes" and "no" logits. This formulation enables the re-ranker to assign higher scores to truly relevant chunks and strengthen its discrimination ability across general and entity-specific domains.
During training, a supervised contrastive loss on the "yes"/"no" logits guides the re-ranker to better distinguish relevant chunks across general and entity-specific domains.
\begin{algorithm}[t]
\caption{Domain-to-Entity Re-ranker Fine-Tuning}
\label{alg:reranker_training}
\KwIn{
    Human-annotated general dataset $\mathcal{D}_{\text{human}} = \{(q, p^+, p^-), \dots\}$ \\
    where $q$ is a query, $p^+$ is a relevant chunk (positive), and $p^-$ is an irrelevant chunk (negative); \\
    Target company queries $\mathcal{Q}_{\text{target}}$ and corpus $\mathcal{C}_{\text{target}}$; \\
    Base re-ranker model $\mathcal{M}_{\text{base}}$, Retriever $\mathcal{R}$, LLM-Annotator $\mathcal{L}_{\text{anno}}$;
}
\KwOut{Specialized re-ranker model $\mathcal{M}_{\text{specialized}}$}
\BlankLine

\SetKwFunction{FMain}{TrainContrastive}
\SetKwProg{Fn}{Function}{:}{}
\Fn{\FMain{$\mathcal{M}$, $\mathcal{D}_{\text{train}}$}}{
    \ForEach{training step}{
        Sample a query $q$ from $\mathcal{D}_{\text{train}}$\;
        Let $\mathcal{P}_q, \mathcal{N}_q$ be the positive and negative sets for $q$\;
        Sample one positive chunk $p^+ \in \mathcal{P}_q$\;
        Sample $K$ negative chunks $\mathcal{N}^- \subset \mathcal{N}_q$\;
        Compute logits $\mathbf{s} = [\text{Score}(q, p^+), \dots, \text{Score}(q, p^-_K)]$ with model $\mathcal{M}$\;
        Compute loss $\mathcal{L} = -\log\left(\frac{e^{s^+}}{e^{s^+} + \sum_{j=1}^{K} e^{s^-_j}}\right)$\;
        Update $\mathcal{M}$ by minimizing $\mathcal{L}$\;
    }
    \Return{$\mathcal{M}$}
}

\BlankLine
\lcc{Stage 1: Financial Re-ranker Training.}\;
$\mathcal{D}_{\text{aug}} \gets \emptyset$\;
\ForEach{$(q, p^+, p^-) \in \mathcal{D}_{\text{human}}$}{
    $q_{\text{aug}}, p^+_{\text{aug}}, p^-_{\text{aug}} \gets \text{ApplyEntityAbstraction}(q, p^+, p^-)$\;
    Add $(q_{\text{aug}}, p^+_{\text{aug}}, p^-_{\text{aug}})$ to $\mathcal{D}_{\text{aug}}$\;
}
$\mathcal{M}_{\text{general}} \gets$ \FMain{$\mathcal{M}_{\text{base}}$, $\mathcal{D}_{\text{aug}}$}\;

\BlankLine
\lcc{Stage 2: Target Company Adaptation}\;
$\mathcal{D}_{\text{auto}} \gets \emptyset$\;
\ForEach{$q \in \mathcal{Q}_{\text{target}}$}{
    $\mathcal{C}_q \gets \mathcal{R}(q, \mathcal{C}_{\text{target}})$ \;%tcp*{Retrieve candidate chunks}
    $\mathcal{P}_q, \mathcal{N}_{\text{hard}} \gets \mathcal{L}_{\text{anno}}(q, \mathcal{C}_q)$ \tcp*{hard negatives}
    $\mathcal{N}_{\text{rand}} \gets \text{SampleRandomChunks}(\mathcal{C}_{\text{target}})$ \tcp*{random negatives}
    $\mathcal{N}_q \gets \mathcal{N}_{\text{hard}} \cup \mathcal{N}_{\text{rand}}$ \;
    Add $(q, \mathcal{P}_q, \mathcal{N}_q)$ to $\mathcal{D}_{\text{auto}}$\;
}
$\mathcal{M}_{\text{specialized}} \gets$ \FMain{$\mathcal{M}_{\text{general}}$, $\mathcal{D}_{\text{auto}}$}\;

\BlankLine
\Return{$\mathcal{M}_{\text{specialized}}$}
\end{algorithm}

\subsubsection{Domain-to-Entity Adaptation Strategy} \label{app:fin_rr_train}
We propose a two-stage, domain-to-entity fine-tuning strategy to develop a re-ranker that is both financially knowledgeable and rapidly adaptable to specific entities. The first stage trains a general financial re-ranker on an abstracted, entity-agnostic dataset, while the second stage specializes the model for a target company using automatically annotated data. This tiered approach maximizes base-model generalization while minimizing manual customization effort.

% \paragraph{Stage 1: Training the General Re-ranker with Data Augmentation.}
% The objective of the first stage is to train a re-ranker that understands the semantic and structural nuances of financial filings, independent of any single company's specific details. To achieve this, we begin with a high-quality, human-expert-annotated dataset of query-chunk pairs. We then apply a data augmentation process to this dataset to improve model generalization and prevent overfitting to specific entity names.
% This abstraction process is implemented through a unified framework that uses an LLM to systematically mask entity-specific information. We employ three abstraction strategies:
% \textbf{i)} Replace specific product models and individual names (e.g., executives, board members) with abstract placeholders. % (e.g., "Product A," "Person B").
% \textbf{ii)} Replace the name of the target company with a placeholder while replacing other company names with randomly sampled competitors. 
% \textbf{iii)} Apply full entity abstraction, replacing all company, product, and individual mentions with consistent placeholders.
% By training on these augmented corpora, the re-ranker is forced to learn financial reasoning detached from memorized facts, which results in a robust, general-purpose model. The optimal abstraction strategy is selected based on comparative performance. More details are listed in Appendix ~\ref{app:data_abstraction}.

\paragraph{\textbf{Stage 1:} Finance Reranker Training.}
The first stage aims to train a re-ranker that captures the semantic and structural nuances of financial filings. Using a high-quality, expert-annotated dataset, we apply data augmentation and systematically mask entity-specific information via three abstraction strategies:
i) Replace product models and individual names (e.g., executives, board members) with placeholders. 
ii) Replace the target company with a placeholder and other companies with randomly sampled competitors. iii) Apply full abstraction, replacing all company, product, and individual mentions with consistent placeholders. Training on these augmented corpora enables the re-ranker to perform financial reasoning independent of memorized facts, yielding a robust, general-purpose model. The optimal abstraction strategy is selected based on comparative performance.

% \paragraph{Stage 2: Rapid Onboarding with Automated Data Annotation.}
% The second stage specializes the general re-ranker for a target company. To simulate a fast and scalable onboarding process, we eliminate the bottleneck of manual data creation by using an automated data annotation module. This is crucial, as any change to the retrieval pipeline would otherwise necessitate a complete and costly re-annotation effort by human experts.
% This module uses an LLM annotator to label retrieved chunks as relevant or irrelevant for a given query, automatically generating the positive and negative pairs needed for fine-tuning. The process operates on the actual output of our Multi-path Retrieval pipeline, ensuring the training data distribution mirrors the inference-time conditions.
% The annotator's judgment is guided by a sophisticated prompt that includes few-shot examples and defines clear relevance criteria (e.g., direct answer, contextual support, fuzzy time match). This automated approach ensures the rapid, consistent, and scalable generation of high-quality training data for specializing the re-ranker to any new company. Further details on the prompt are available in Appendix~\ref{app:annotation_prompt}.

\paragraph{\textbf{Stage 2:} Target Company Adaptation.}
The second stage specializes the general re-ranker for a target company. To eliminate the manual data creation bottleneck, we use an automated annotation module where an LLM labels retrieved chunks as relevant or irrelevant. 
This process operates on the actual output of our Multi-path Retrieval pipeline, ensuring the training data distribution mirrors the inference-time conditions for robust model performance.
The annotator's judgment is guided by a sophisticated prompt that includes few-shot examples and defines clear relevance criteria (e.g., direct answer, contextual support, fuzzy time match). This automated approach ensures the rapid, consistent, and scalable generation of high-quality training data for specializing the re-ranker to any new company. Further details on the annotation prompt are in Appendix~\ref{app:annotation_prompt}.

\subsubsection{Contrastive Learning Objective}  
Both fine-tuning stages use supervised fine-tuning (SFT) with a contrastive learning objective. For a query $q$, the model is trained to assign higher scores to positive (relevant) chunks $p^+$ than to $K$ negative (irrelevant) chunks $\{p^-_j\}_{j=1}^K$.  

For training, all chunks retrieved for $q$ are annotated: relevant chunks form the positive set $\mathcal{P}_q = \{p^+_1, \dots, p^+_n\}$, and irrelevant chunks form the negative set $\mathcal{N}_q = \{p^-_1, \dots, p^-_m\}$. To enhance robustness, $\mathcal{N}_q$ combines (1) hard negatives from retriever-identified but irrelevant chunks, and (2) random negatives sampled from the corpus \citep{Huang_2020}. Each training quadruple is structured as
\begin{equation}
\mathcal{D} = \big\{\langle q, \mathcal{P}_q, \mathcal{N}_q, \text{prompt} \rangle\big\}.
\end{equation} 
For each query, a mini-batch contains one positive $p^+ \in \mathcal{P}_q$ and $K$ negatives $\{p^-_1, \dots, p^-_K\} \subset \mathcal{N}_q$, producing scores $\mathbf{s} = [s^+, s^-_1, \dots, s^-_K]$ with $s_i = \text{Score}(q, p_i)$.  

The $(K+1)$ candidates are treated as an \textit{ad-hoc} set of classes, converting the task into classification where $p^+$ is the positive class. The training loss is standard cross-entropy:
\begin{equation}
\mathcal{L} = -\log \frac{e^{s^+}}{e^{s^+} + \sum_{j=1}^{K} e^{s^-_j}}.
\end{equation}
This objective encourages the model to maximize $s^+$ relative to all negatives, learning to distinguish relevant from irrelevant content.  

For Stage 1, negatives in the human-annotated dataset $\mathcal{D}_{\text{human}}$ serve as hard negatives. Stage 2 uses hard negatives generated by our LLM-Annotator. Training is implemented with FlagEmbedding\footnote{\url{https://github.com/FlagOpen/FlagEmbedding}} to fine-tune \texttt{bge-reranker-v2-gemma} \cite{chen2024bgem3embeddingmultilingualmultifunctionality}. Full procedure and hyperparameters are in Procedure~\ref{alg:reranker_training} and Appendix~\ref{app:reranker_training}.

% \subsubsection{Contrastive Learning Objective}
% Both fine-tuning stages are driven by supervised fine-tuning (SFT) on a contrastive learning objective. The goal is to train the model to assign a higher score to a positive (relevant) chunk $p^+$ than to set of $K$ negative (irrelevant) chunks  $\{p_j^-\}_{j=1}^K$ for a given query $q$. The training objective is the standard cross-entropy loss over the scores of the positive and negative samples. For a mini-batch containing one positive sample $p^+$ and $K$ negative samples, the loss is:
% \begin{equation}
% \mathcal{L} = -\log\left(\frac{e^{s^+}}{e^{s^+} + \sum_{j=1}^{K} e^{s^-_j}}\right)
% \end{equation}
% where is the re-ranker's score 

% \subsubsection{Model Evaluation}
% Requires: \usepackage{booktabs, makecell}

\subsection{Memory Bank Look-up} \label{sub:high-frequency-QA}
The Memory Bank Look-up module provides instantaneous answer to common and recurring financial queries with the pre-trained Memory Bank (Sec. \ref{subsub: high-frequency-memory-bank}). After matching the incoming user query with the pre-compiled knowledge base verified by human expert, the system can bypass computationally intensive deep retrieval, reasoning and generation steps, hence reducing overall latency. We normalize the query to remove punctuation and irrelevant entities, and match the query to knowledge base with the three parallel algorithms: 
(1) \textbf{Subsequence-based Similarity} computes structural similarity between user's query and canonical questions. It identifies the longest contiguous matching subsequences of characters between the two strings to capture precise terminology and word order (sub-sequence). 
% The score $S_{seq}$ is calculated by \begin{equation}
% S_{\mathrm{seq}}\left(q, c_k\right)=\frac{2 \cdot \operatorname{LCSM}\left(q, c_k\right)}{|q|+\left|c_k\right|}
% \end{equation}, where $\text{LCSM}(q, c_k)$ is the length of the longest contiguous matching subsequence between query and canonical question, and $|s|$ denotes the length of the string $s$.
(2) \textbf{Keyword-based Similarity (BM25)} handles variations in phrasing by applying the BM25 algorithm on structure table. This method measures the relevancy of a canonical question based on keyword overlap with the user's query, using the scoring function discussed in Section \ref{sub:multi-path-retrieval}.
(3) \textbf{Semantic Similarity} tackles queries that are syntactically different but semantically equivalent as the canonical questions. We utilize the dense retrieval approach mentioned in Section \ref{sub:multi-path-retrieval} to identify a match based on conceptual intent instead of lexical overlap.

We consider a canonical question $c_{k}$ potential match if its score from any of the three methods $S_{seq}(q,c_{k})$, $S_{bm25}(q,c_{k})$, $S_{sem}(q,c_{k})$ surpasses the predefined confidence thresholds $\tau_{\text{seq}}$, $\tau_{\text{bm25}}$, and $\tau_{\text{sem}}$. The final matching decision for a given pair $(q,c_{k})$ is determined by the boolean function $M(q, c_{k})$: 
% \begin{equation}
% \begin{split}
% M(q, c_k) ={}& (S_{\text{seq}}(q, c_k) > \tau_{\text{seq}}) \\
%               & \vee (S_{\text{bm25}}(q, c_k) > \tau_{\text{bm25}}) \\
%               & \vee (S_{\text{sem}}(q, c_k) > \tau_{\text{sem}})
% \end{split}
% \end{equation}

\begin{equation}
\begin{split}
M(q, c_k) ={}& (S_{\text{seq}}(q, c_k) > \tau_{\text{seq}})  \vee (S_{\text{bm25}}(q, c_k) > \tau_{\text{bm25}}) \\
              & \vee (S_{\text{sem}}(q, c_k) > \tau_{\text{sem}}).
\end{split}
\end{equation}
If $M(q, c_{k})$ is true for any $c_{k}$, the system retrieves canonical question $c_{k}$ and the corresponding entry from memory bank to return the answr in a structured key-value format. 

\subsection{Tool Use} \label{sub:tool-call}
% WHW
While retrieval from SEC filings covers most historical data, queries requiring real-time information, such as current stock prices or recent corporate actions, cannot be addressed without external sources. The Tool Use module addresses this gap by integrating external APIs to fetch current market information, operating asynchronously alongside other retrieval streams, and contributing real-time context that complements the historical analysis with real-time context.
The module adheres to OpenAI's function-calling schema, enabling the LLM to autonomously select and invoke appropriate tools based on query requirements. Currently supported tools include real-time market data retrieval and event-specific information access (e.g., IPO details) via a third-party financial API. Upon execution, tool outputs are integrated into the conversational context as additional evidence for final answer synthesis.
% The standardized implementation provides flexibility and extensibility for future system enhancement: while the current toolset focuses on financial data, new capabilities can be seamlessly integrated by defining their JSON schemas and implementing the corresponding API functions.

\vspace{-0.3cm}
\section{Experiments}
\subsection{Datasets and Baseline}

% \paragraph{Datasets and Baselines}
We evaluate \ourmodel{} across 4 financial QA datasets: \textbf{FinanceBench} \citep{islam2023financebench}, containing human-verified QA pairs with single gold reference chunks from financial filings; \textbf{FinQA} \citep{chen2022finqadatasetnumericalreasoning}, from which we use the textual reasoning subset with single reference chunks; and 2 \textbf{In-House Company Dataset} with multi-entity (e.g., Lotus, Zeekr) QA pairs and annotated relevance labels. We compare our model against graph-based RAG systems including \textbf{GraphRAG} and \textbf{LightRAG} \citep{guo2025lightragsimplefastretrievalaugmented}, as well as standard RAG baselines that perform direct retrieval-to-generation.

\begin{table*}[t]
\caption{\textbf{End-to-End Performance (top-10).} Four panels (2×2). CAKC = Context-Aware Knowledge Curation; Correct. = Correctness; Resp. Rel. = Response Relevancy; Ctx. Recall/Prec. = Context Recall/Precision. \emph{Avg.} is the unweighted mean of \{Correct., Resp. Rel., Ctx. Recall, Ctx. Prec.\}. Values are \%. Our method is in \textbf{bold}.}
\label{tab:e2e_2x2}
\vspace{-0.4cm}
\centering
\small
\setlength{\tabcolsep}{3pt}
\renewcommand{\arraystretch}{0.9}

% Inner header macro
\newcommand{\ablinnerheader}{%
\toprule
\thead{w/ CAKC} & \thead{Config} & \thead{Correct.} &
\thead{Resp.\\Rel.} & \thead{Ctx.\\Recall} & \thead{Ctx.\\Prec.} & \thead{Avg.} \\
\midrule
}

% ===== Row 1: Zeekr | Lotus =====
\begin{minipage}[t]{0.49\textwidth}
\centering
\textbf{Zeekr}\\
\begin{tabular}{@{}llccccc@{}}
\ablinnerheader
No  & GraphRAG        & 23.3 & 38.9 & 0.1  & 11.7 & 18.5 \\
No  & LightRAG        & 20.5 & 46.8 & 11.2 & 19.3 & 24.5 \\
No  & BM25            & 6.0  & 16.2 & 3.8  & 27.2 & 13.3 \\
No  & Faiss           & 3.8  & 12.0 & 2.3  & 15.7 & 8.4  \\
No  & BM25+Faiss      & 5.6  & 15.0 & 3.1  & 23.2 & 11.7 \\
% \cmidrule(lr){1-7}
\midrule
Yes & BM25            & 29.4 & 63.0 & 30.3 & 44.7 & 41.9 \\
Yes & Faiss           & 37.4 & 69.6 & 36.5 & 41.0 & 46.1 \\
Yes & BM25+Faiss      & 35.9 & 69.5 & 36.0 & 41.1 & 45.6 \\
Yes & \textbf{\ourmodelb{} (ours)} & \textbf{64.3} & \textbf{82.0} & \textbf{59.0} & \textbf{46.3} & \textbf{62.9} \\
\bottomrule
\end{tabular}
\end{minipage}\hfill
\begin{minipage}[t]{0.49\textwidth}
\centering
\textbf{Lotus}\\
\begin{tabular}{@{}llccccc@{}}
\ablinnerheader
No  & GraphRAG        & 25.9 & 43.8 & 0.4  & 8.2  & 19.6 \\
No  & LightRAG        & 24.8 & 58.8 & 15.9 & 28.7 & 32.1 \\
No  & BM25            & 17.5 & 34.2 & 7.0  & 23.5 & 20.6 \\
No  & Faiss           & 16.0 & 31.3 & 6.5  & 20.6 & 18.6 \\
No  & BM25+Faiss      & 16.0 & 22.4 & 6.4  & 23.7 & 17.1 \\
% \cmidrule(lr){1-7}
\midrule
Yes & BM25            & 39.6 & 79.0 & 37.2 & 44.0 & 50.0 \\
Yes & Faiss           & 39.7 & 81.4 & 45.5 & 40.4 & 51.8 \\
Yes & BM25+Faiss      & 46.5 & 77.9 & 48.2 & 46.8 & 54.9 \\
Yes & \textbf{\ourmodelb{} (ours)} & \textbf{57.3} & \textbf{89.4} & \textbf{57.6} & \textbf{51.6} & \textbf{64.0} \\
\bottomrule
\end{tabular}
\end{minipage}

\vspace{4pt}

% ===== Row 2: FinanceBench | FinQA Bench =====
\begin{minipage}[t]{0.49\textwidth}
\centering
\textbf{FinanceBench}\\
\begin{tabular}{@{}llccccc@{}}
\ablinnerheader
No  & GraphRAG        & 9.2           & 6.8  & 2.6  & 2.7           & 5.33 \\
No  & LightRAG        & \textbf{40.3} & 6.5  & 3.1  & 8.8           & 14.7 \\
No  & BM25            & 17.7          & 54.3 & 34.6 & 39.8          & 36.6 \\
No  & Faiss           & 13.7          & 51.3 & 30.4 & 40.6          & 34.0 \\
No  & BM25+Faiss      & 18.6          & 53.4 & 32.5 & \textbf{46.8} & 37.8 \\
% \cmidrule(lr){1-7}
\midrule
Yes & BM25            & 22.8 & 32.6 & 29.4          & 15.3 & 25.0 \\
Yes & Faiss           & 20.9 & 63.2 & \textbf{50.8} & 35.9 & 42.7 \\
Yes & BM25+Faiss      & 21.7 & 52.8 & 46.0          & 28.1 & 37.1 \\
Yes & \textbf{\ourmodelb{} (ours)} & 24.7 & \textbf{75.1} & 49.2 & 37.2 & \textbf{46.5} \\
\bottomrule
\end{tabular}
\end{minipage}\hfill
\begin{minipage}[t]{0.49\textwidth}
\centering
\textbf{FinQA Bench}\\
\begin{tabular}{@{}llccccc@{}}
\ablinnerheader
No  & GraphRAG        & 22.8 & 53.1 & 4.6  & 2.1            & 25.8 \\
No  & LightRAG        & 29.5 & 64.1 & 34.0 & 33.4           & 40.3 \\
No  & BM25            & 30.3 & 54.0 & 32.4 & 40.8           & 39.4 \\
No  & Faiss           & 28.2 & 51.1 & 29.1 & 39.6           & 37.0 \\
No  & BM25+Faiss      & 30.9 & 53.4 & 32.1 & \textbf{45.2}  & 40.4 \\
% \cmidrule(lr){1-7}
\midrule
Yes & BM25            & 31.4 & 72.6 & 73.3          & 27.8 & 51.3 \\
Yes & Faiss           & 33.4 & 76.8 & \textbf{75.7} & 29.1 & 53.7 \\
Yes & BM25+Faiss      & 32.0 & 74.8 & 74.7          & 29.6 & 52.8 \\
Yes & \textbf{\ourmodelb{} (ours)} & \textbf{33.7} & \textbf{80.1} & 73.9 & 38.8 & \textbf{56.6} \\
\bottomrule
\end{tabular}
\end{minipage}

\end{table*}

\subsection{Experiment Settings}
\subsubsection{Multi-path Retrieval (MPR) settings}
To evaluate the efficiency of our Multi-path Retrieval module, we perform retrieval tasks on both in-house company dataset and open-source dataset Financebench \citep{islam2023financebench}. Preprocessing for this experiment consists of three stages: (1) segmenting financial filings into chunks using CAKC, (2) indexing the chunks in a vector store, and (3) rewriting user queries into a format optimized for RAG systems. To ensure a fair comparison across different retriever configurations, we adjust the top-k parameter for each setting to maintain a comparable total number of unique retrieved chunks. Retrieval quality is measured by the evidence hit rate, where a retrieved chunk is classified as a hit if its cosine similarity to the ground-truth evidence exceeds a predefined threshold. We use Qwen/Qwen-Embedding-4B \citep{qwen3technicalreport} as our embedding model. For in-house dataset, this threshold is set to 1.0, as the ground-truth evidence chunks are the exact chunks selected from the vector store.

\begin{table*}[ht]
    \caption{
        Reranker performance on company dataset under different retrieval settings (SFT with Zeekr + Lotus Data).\\
        Green gains next to Stage 1 indicate improvements over the pretrained baseline. 
        Dark-green gains next to S2 (Stage 2) indicate further improvements over the best S1 (Stage 1: Complete Abstraction) variant.
    }
    \small
    \centering
    \setlength{\tabcolsep}{3pt}
    % \renewcommand{\arraystretch}{1.3}

    % color for stage2 (deeper green)
    \resizebox{\linewidth}{!}{%
        \begin{tabular}{
            p{1.2cm} | 
            p{1.6cm} | 
            l | 
            c c c c | 
            c c c c 
        }
            \toprule
            \textbf{Dataset} & \textbf{Base} & \textbf{Training Data} 
            & \textbf{NDCG@5} & \textbf{MRR@5} & \textbf{Precision@5} & \textbf{Recall@5} 
            & \textbf{NDCG@10} & \textbf{MRR@10} & \textbf{Precision@10} & \textbf{Recall@10} \\
            % \midrule\midrule
            % \specialrule{\heavyrulewidth}{0pt}{0pt}
            \midrule
            % ----------------------- LOTUS -----------------------
            \multirow{7.5}{*}{\centering\bfseries Lotus}
                & \bfseries PT & N/A (Baseline) 
                & 62.0 & 60.1 & 29.3 & 23.0 
                & 67.3 & 68.2 & 24.4 & 34.8 \\
                \cmidrule{2-11}
                & \multirow{3}{*}{\bfseries \shortstack{PT+S1}}
                    & w/ Product/Person Abs. D. 
                    & 64.2 \gains{2.2} & 62.7 \gains{2.6} & 32.2 \gains{2.9} & 25.8 \gains{2.8}
                    & 69.3 \gains{2.0} & 69.0 \gains{0.8} & 26.7 \gains{2.3} & \underline{38.5 \gains{3.7}} \\
                    &
                    & w/ Company Name Abs. D.
                    & 65.1 \gains{3.1} & 65.2 \gains{5.1} & 31.1 \gains{1.8} & 24.2 \gains{1.2}
                    & \underline{69.5 \gains{2.2}} & \underline{71.5 \gains{3.3}} & 25.6 \gains{1.2} & 37.3 \gains{2.5} \\
                    &
                    & w/ Complete Abs. Dataset
                    & \underline{69.2 \gains{7.2}} & \underline{69.1 \gains{9.0}} & \underline{33.4 \gains{4.1}} & \underline{26.1 \gains{3.1}}
                    & 69.2 \gains{1.9} & 71.0 \gains{2.8} & \underline{27.1 \gains{2.7}} & 38.4 \gains{3.6} \\
                \cmidrule{2-11}
                & \bfseries PT+S2
                & Company Specific Dataset
                & 72.6 \gainsdeep{3.4} & 71.5 \gainsdeep{2.4} & 35.3 \gainsdeep{1.9} & 27.9 \gainsdeep{0.4}
                & 71.5 \gainsdeep{2.3} & 73.4 \gainsdeep{2.4} & 27.5 \gainsdeep{0.4} & 39.0 \gainsdeep{0.6} \\
                \cmidrule{2-11}
                & \bfseries \shortstack{PT+S1+S2} 
                & Company Specific Dataset
                & \textbf{75.1} \gainsdeep{5.9} & \textbf{73.5} \gainsdeep{4.4} & \textbf{37.4} \gainsdeep{4.0} & \textbf{28.3} \gainsdeep{2.2}
                & \textbf{75.4} \gainsdeep{6.2} & \textbf{78.6} \gainsdeep{7.6} & \textbf{28.1} \gainsdeep{1.0} & \textbf{39.5} \gainsdeep{1.1} \\
            \midrule
            % \specialrule{\heavyrulewidth}{0pt}{0pt}
            % ----------------------- ZEEKR -----------------------
            \multirow{7.5}{*}{\bfseries Zeekr}
                & \bfseries PT & N/A (Baseline) 
                & 77.1 & 70.0 & 64.9 & 32.1 
                & 74.2 & 65.3 & 57.4 & 43.1 \\
                \cmidrule{2-11}
                & \multirow{3}{*}{\bfseries \shortstack{PT+S1}}
                    & w/ Product/Person Abs. D. 
                    & 79.8 \gains{2.7} & 73.4 \gains{3.4} & 67.9 \gains{3.0} & 34.2 \gains{2.1}
                    & 74.2 \neutral & 66.3 \gains{1.3} & 59.3 \gains{1.9} & \underline{45.7 \gains{2.6}} \\
                    &
                    & w/ Company Name Abs. D. 
                    & 81.6 \gains{4.5} & 75.1 \gains{5.1} & 69.9 \gains{5.0} & 35.0 \gains{2.9}
                    & 75.7 \gains{1.5} & 67.4 \gains{2.1} & 59.5 \gains{2.1} & 44.6 \gains{1.5} \\
                    &
                    & w/ Complete Abs. Dataset 
                    & \underline{82.6 \gains{5.5}} & \underline{75.9 \gains{5.9}} & \underline{71.5 \gains{6.6}} & \underline{35.7 \gains{3.6}}
                    & \underline{76.3 \gains{2.1}} & \underline{69.4 \gains{4.1}} & \underline{59.9 \gains{2.5}} & 44.4 \gains{1.3} \\
                \cmidrule{2-11}
                & \bfseries PT+S2
                & Company Specific Dataset
                & 84.4 \gainsdeep{1.8} & 77.3 \gainsdeep{1.4} & 75.5 \gainsdeep{4.0} & 38.8 \gainsdeep{3.1}
                & 76.5 \gainsdeep{0.2} & 68.4 \loss{1.0}  & 61.5 \gainsdeep{1.6} & 46.5 \gainsdeep{2.1} \\
                \cmidrule{2-11}
                & \bfseries \shortstack{PT+S1+S2} 
                & Company Specific Dataset 
                & \textbf{87.1} \gainsdeep{4.5} & \textbf{83.6} \gainsdeep{7.7} & \textbf{77.2} \gainsdeep{5.7} & \textbf{39.0} \gainsdeep{3.3}
                & \textbf{78.2} \gainsdeep{1.9} & \textbf{70.9} \gainsdeep{1.5} & \textbf{63.4} \gainsdeep{3.5} & \textbf{47.1} \gainsdeep{2.7} \\
            \bottomrule
        \end{tabular}
    }
    \label{tab:performance_mad_grouped}
\end{table*}
\begin{figure*}[htbp] 
    \centering  
    \includegraphics[width=1\linewidth]{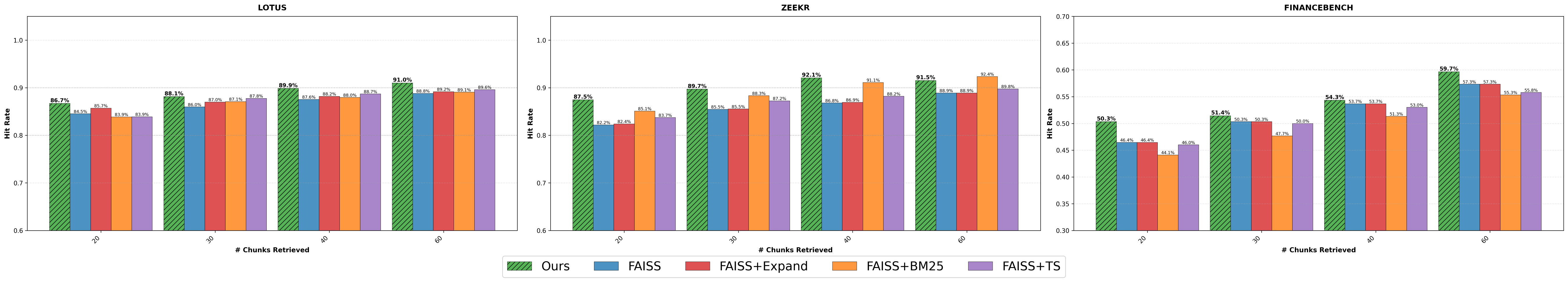}
    \caption{Hit rate comparison across multi-path retrieval (MPR) settings. Our hybrid retrieval approach consistently improves evidence hit rates across evaluation datasets in most retrieval scenarios.}
    \label{fig:retrieval}
\end{figure*}
\subsubsection{Domain-to-Entity Adaptation Re-ranking (DAR) settings}
The re-ranker evaluation is designed to rigorously test two distinct performance objectives: general financial domain competence and target company specialization. %To this end, we fine-tune the LLM-based models BAAI/bge-reranker-v2-gemma (2.5B),  utilizing LoRA \citep{hu2021loralowrankadaptationlarge} for efficiency and a contrastive learning objective for relevance scoring. Training data consists of query-chunk pairs constructed with a mix of hard negatives (retriever failures) and easy random negatives to ensure robust relevance learning, and t
The model performance is evaluated using standard information retrieval metrics at K=5 and K=10: NDCG@K, MRR@K, Precision@K, and Recall@K.
The evaluation is structured around a two-stage benchmarking process, with specific baselines defined for each stage to quantify the performance gain. \textbf{Stage 1} assesses the model's ability to generalize across financial entities. The re-ranker is trained on the union of abstracted dataset from both Lotus and Zeekr. The performance baseline for Stage 1 is the re-ranker's own pretrained version. We evaluate the three abstraction strategies and select the checkpoint with the highest Precision@5 as the optimal general model as well as the base model for Stage 2 fine-tuning. \textbf{Stage 2} assesses the specialization for the target company, in which the baseline is defined by a control experiment: the re-ranker's pretrained version directly fine-tuned with LoRA \citep{hu2021loralowrankadaptationlarge} on the company-specific dataset (the checkpoint that achieved the highest Precision@5 in this control run). This allows us to quantify the superiority of the two-stage approach over immediate specialization.

\subsubsection{End-to-End response settings}
% QA accuracy on Each dataset
% GraphRAG
% LightRAG
% MPR
% MPR + HighFrequencyQA
% MPR + HighFrequencyQA + ToolCall (SageFin)
% We evaluate the end-to-end question answering performance of \ourmodel{} by comparing it against the baselines on the evaluation datasets. To provide a holistic assessment, we employed five distinct metrics from the RAGAS framework \citep{es2025ragasautomatedevaluationretrieval}, which collectively measure the quality of both the retrieval and generation stages. 1) Factual Correctness: Measures the factual accuracy of the generated response by comparing it against a ground-truth reference answer. 2) Faithfulness: Evaluates the factual consistency of the response against the retrieved context. A high faithfulness score indicates the answer is grounded in the provided information and free of hallucination. 3) Response Relevancy:  Assesses the degree to which the generated answer directly addresses the user's original question. 4) Context Recall: Measures the extent to which the retrieved context contains all the information necessary to answer the question, based on the ground-truth answer. 5) Context Precision: evaluates the signal-to-noise ratio of the retrieved context, assessing whether the most relevant chunks are ranked at the top.
We evaluate the end-to-end question-answering performance of \ourmodel{} against baselines using the evaluation datasets. Settings for GraphRAG and LightRAG are included in (Appendix. \ref{sub: graphrag-lightrag-config}).
To provide a holistic assessment of both retrieval and generation quality, we adopt four metrics from the RAGAS framework \citep{es2025ragasautomatedevaluationretrieval}: Factual Correctness, which measures the factual alignment of the generated response with the reference answer; Response Relevancy, which assesses how directly the response addresses the user query; Context Recall, which quantifies the coverage of relevant information in the retrieved context; and Context Precision, which evaluates the proportion of relevant information among the retrieved passages. These metrics together capture the overall faithfulness and effectiveness of the RAG pipeline, and are detailed in Appendix~\ref{app:ragas}.
\section{Results and Analysis}

\subsection{End-to-end Response Results}
We evaluate the end-to-end question-answering performance of \textbf{\ourmodelb} against a suite of baselines using four datasets: two proprietary in-house knowledge bases (Zeekr, Lotus) and two public financial benchmarks (FinanceBench \citep{islam2023financebench} and FinQA Bench \citep{chen2022finqadatasetnumericalreasoning}). The comprehensive results, summarized in Table \ref{tab:e2e_2x2}, demonstrate the consistent superiority of our method as it achieves the highest unweighted mean (\emph{Avg.}) performance across all four datasets. A key observation is that the inclusion of Context-Aware Knowledge Curation (CAKC) provides a substantial performance boost to all retrieval methods, resulting in noticeable gains across metrics, particularly in Context Recall and Factual Correctness, by enabling better integration of structured and context-comprehensive knowledge during the generation phase. Specifically, \textbf{\ourmodelb} leads the performance on the proprietary datasets, demonstrating superior \textbf{Factual Correctness} and \textbf{Response Relevancy} on Zeekr and Lotus, respectively, confirming that the responses are highly accurate and well-aligned with user intent. Collectively, this strong performance validates the effectiveness of our integrated architecture in achieving superior and generalizable end-to-end question-answering capability.
\subsection{Multi-path Retrieval Results}
As shown in Figure~\ref{fig:retrieval}, our multi-path retrieval approach consistently outperforms all baseline methods across both in-house datasets (Lotus, Zeekr) and open-source (Financebench) datasets. While the Faiss-only method achieves competitive performance, retrieving most relevant evidence within the top 20 chunks, our ensemble approach combining four retrieval methods introduces greater diversity and yields 2-5\% improvement in hit rate across varying chunk counts. Notably, we observe dataset-dependent performance characteristics among individual methods: Faiss+Bm25 demonstrates strong performance on Zeekr but shows degraded results on Financebench, suggesting different underlying data distributions and query patterns. This variability underscores the robustness of our multi-path ensemble, which maintains superior and consistent performance across all datasets by leveraging complementary strengths of different retrieval strategies.

\subsection{Domain-to-entity Re-ranking Results}
Table \ref{tab:performance_mad_grouped} shows that our two-stage Domain-to-Entity adaptation strategy (See Section \ref{app:fin_rr_train}) achieves consistent and substantial improvements across all metrics on both the Lotus and Zeekr datasets. Stage 1 delivers notable gains over the pretrained baseline, demonstrating the value of abstracted, entity-agnostic training for developing a robust financial re-ranker. Among the variants, the model trained on the Complete abstract dataset performs the best, indicating that fully masking entity mentions helps the model focus on financial semantics rather than surface-level patterns. Building on this foundation, Stage 2 further enhances performance, surpassing both Stage 1 and Direct SFT by effectively refining company-specific relevance estimation. Compared with Direct SFT, which fine-tunes solely on company-specific data, Model after DAR achieves higher accuracy on both datasets, yielding boosts of over 2.5 in NDCG@5 and 2.0 in MRR@5 on Zeekr, and similar gains on Lotus. These results demonstrate that Stage 1 establishes a transferable understanding of financial language, while Stage 2 tailors this knowledge to new entities through automated, retrieval-aligned adaptation.
\subsection{System Latency and Cost Results}
We conduct a comprehensive end-to-end analysis of the system's operational efficiency, measuring both \textbf{system latency} and \textbf{deployment cost}. This evaluation assesses the performance of every module in our processing pipeline, specifically quantifying the average execution time and the token usage for each large language model (LLM) invocation. Our analysis particularly focuses on the variable steps, such as \textbf{Sub-query Answering} and \textbf{Tool Use}, to understand how system performance scales with increasing query complexity and tool integration. For a detailed breakdown of the time and cost associated with each processing step, please refer to Appendix ~\ref{app:cost_analysis}.
\section{Conclusion}
\vspace{+0.2cm}

This paper introduced \ourmodel{}, a practical end-to-end framework for financial QA that addresses the dual challenge of processing heterogeneous disclosures and achieving efficient company-specific adaption. Through the synergy of the \emph{Context-Aware Knowledge Curation}, \emph{Tripartite Hybrid Retrieval} engine and \emph{two-stage Domain-to-Entity re-ranking} training strategy, \ourmodel{} delivers strong end-to-end performance, with the largest improvement reaching \textbf{+38.4} points in average on Zeekr compared to LightRAG \citep{guo2025lightragsimplefastretrievalaugmented}. Our ablation studies confirm that the three components provide complementary strengths, forming a cohesive and scalable architecture for grounded financial QA.
% Ablation study indicate that CAKC improves evidence quality and recall, THR increases robustness via complementary retrieval signals, and the two-stage re-ranker produces reliable entity-specific gains with modest supervision. Together, these components form a scalable recipe for accurate, grounded financial QA.
Beyond these empirical gains, \ourmodel{} establishes a generalizable architectural blueprint for domain-specific RAG systems that must balance broad applicability with rapid specialization—a requirement increasingly critical across sectors beyond finance. Future work could explore extending this framework to other specialized domains (e.g., legal, healthcare) and incorporating advanced reasoning capabilities to handle complex multi-hop queries over temporal financial data.

% By unifies \emph{Context-Aware Knowledge Curation} (CAKC), a \emph{Tripartite Hybrid Retrieval} (THR) engine, and a \emph{two-stage Domain-to-Entity re-ranking} strategy.
 % By jointly addressing heterogeneous disclosures (text, tables, figures), retrieval coverage and freshness (multi-path + tool use + high-frequency memory), and rapid company-level adaptation,

% \begin{acks}
% \end{acks}

\bibliographystyle{ACM-Reference-Format}
\bibliography{citations, software}

\appendix
\section{Related Work}
\subsection{Retrieval Augmented Generation}
Large Language Models have demonstrated remarkable capabilities in natural language understanding and generation; however, they face significant challenges in knowledge-intensive tasks, primarily due to their knowledge cutoff and a propensity for hallucination \citep{Huang_2025}. To address this, the initial RAG framework \citep{lewis2021retrievalaugmentedgenerationknowledgeintensivenlp}, built upon the Dense Passage Retrieval (DPR) model \citep{karpukhin2020densepassageretrievalopendomain}, was introduced to augment LLM responses with relevant document chunks from an external knowledge database. While earlier models like DPR focused on extractive tasks, RAG's key innovation was extending this retrieval mechanism to generative tasks, enabling the model to produce coherent and factually grounded responses.

Modern RAG systems have evolved from simple retriever–generator architectures into sophisticated multistage pipelines comprising pre-retrieval and post-retrieval processes \citep{gao2024retrievalaugmentedgenerationlargelanguage}. The pre-retrieval stage focuses on improving query formulation through techniques such as query rewriting \citep{ma2023queryrewritingretrievalaugmentedlarge} and decomposition, while the post-retrieval stage commonly employs re-ranking to refine the candidate set. Traditional cross-encoder re-rankers \citep{nogueira2020passagererankingbert} jointly encode query–document pairs for fine-grained scoring, whereas LLM-based re-rankers \citep{zhang2025qwen3embeddingadvancingtext} leverage generative reasoning to handle more complex ranking scenarios.

Despite recent advances, RAG systems that rely solely on unstructured text remain limited in capturing complex, multi-hop relationships between entities. To address this, GraphRAG and LightRAG frameworks \citep{edge2025localglobalgraphrag, guo2025lightragsimplefastretrievalaugmented, hu2025graggraphretrievalaugmentedgeneration} integrate structured knowledge graphs into retrieval to model inter-entity dependencies and improve factual grounding. While these methods enhance reasoning and coverage, they require expensive graph construction, indexing, and maintenance, making them time-consuming and difficult to scale in dynamic domains.

\subsection{RAG in Finance} 
These limitations are particularly pronounced in the financial domain, where precision and timeliness are critical for regulatory compliance and investment decision-making \citep{Gozman2014-ou}. Financial documents contain dense terminology, rapidly evolving information, and heterogeneous formats such as tables and reports, making retrieval especially challenging \citep{choi2025finderfinancialdatasetquestion, zhao2025finragbenchvbenchmarkmultimodalrag}. To address these challenges, prior work explores three main directions. Some methods focus on internal document information, leveraging structural cues or entity co-occurrence to enhance contextual retrieval \citep{10.1145/2797115.2797120, chatterjee2024dreqdocumentrerankingusing}. Others incorporate external knowledge, such as financial databases or knowledge graphs, to improve factual grounding and reduce hallucination \citep{chen-etal-2024-knowledge}. More recently, hybrid RAG systems attempt to combine both internal and external signals through complex multi-stage pipelines \citep{kim2025optimizingretrievalstrategiesfinancial, lee2024multirerankermaximizingperformanceretrievalaugmented}. While effective, these approaches often require extensive graph construction, corpus refinement, and fine-tuning, leading to high computational cost and limited scalability across financial entities and domains.

\subsection{Financial RAG System Evaluation} The systematic evaluation of financial RAG systems is supported by a growing number of specialized benchmarks, largely centered on question answering from SEC filings. A core set of datasets, including Financebench \citep{islam2023financebench}, FinQA \citep{chen2022finqadatasetnumericalreasoning}, FinDER \citep{choi2025finderfinancialdatasetquestion}, and FinBen \citep{xie2024finbenholisticfinancialbenchmark}, provide (question, context, answer) tuples with golden evidence, enabling assessment of both retrieval quality and end-to-end accuracy. To address the crucial need for quantitative analysis, benchmarks such as FinQA \citep{chen2022finqadatasetnumericalreasoning}, ConvFinQA \citep{chen2022convfinqaexploringchainnumerical}, TAT-QA \citep{zhu2021tatqaquestionansweringbenchmark}, and DocFinQA \citep{reddy2024docfinqalongcontextfinancialreasoning} specifically target numerical reasoning over combined textual and tabular data. Further advancing the rigor of evaluation, recent benchmarks introduce greater complexity. T²RagBench \citep{zhang2025t2rbenchbenchmarkgeneratingarticlelevel} uses re-created, context-independent questions to test true retrieval-based reasoning, while SECQUE \citep{yoash2025secquebenchmarkevaluatingrealworld} raises the difficulty with cross-company questions that require synthesis of information across long contexts.

Manually evaluating the outputs of a RAG system is impractical at scale, making automated evaluation essential. Traditional metrics like ROUGE \citep{lin-2004-rouge} and EM (Exact Match) are often insufficient as they rely on lexical overlap and fail to capture semantic correctness. Consequently, the field has increasingly adopted powerful LLMs as judges. This approach takes two common forms: RAGAS\citep{es2025ragasautomatedevaluationretrieval} and FinDER \citep{choi2025finderfinancialdatasetquestion} prompt an LLM to evaluate a response based on a criteria tuple (faithfulness, correctness, context relevance). Alternatively, as seen in the SECQUE benchmark \citep{yoash2025secquebenchmarkevaluatingrealworld}, an LLM can be used to map a response directly to a holistic quality score.
\section{Re-ranker Fine-tuning} \label{app:reranker_training}
We employ Low-Rank Adaptation (LoRA) \citep{hu2021loralowrankadaptationlarge} through the \textbf{FlagEmbedding} framework to reduce the number of trainable parameters and enhance training efficiency. Specifically, LoRA was enabled with the rank set to 32 and the scaling factor ($\alpha$) set to 64, targeting the attention projection layers ($q_\text{proj}$, $k_\text{proj}$, $v_\text{proj}$, $o_\text{proj}$). The model was fine-tuned for 100 epochs. The learning rate was set to 1e-4 to ensure stable convergence. To manage memory and maintain a small effective batch size, the per-device batch size was set to 1, and gradient accumulation was enabled with \textit{gradient\_accumulation\_steps}=2. Mixed precision training was employed using the \textit{bf16} setting to further optimize memory usage. The training utilized a warmup ratio of 0.1 and a weight decay of 0.1. Input sequences were configured with a \textit{train\_group\_size} of 8.

\section{High Frequency Memory Bank} \label{app:freq_table}
\begin{figure}[h] 
    \centering 
    \includegraphics[width=1\linewidth]{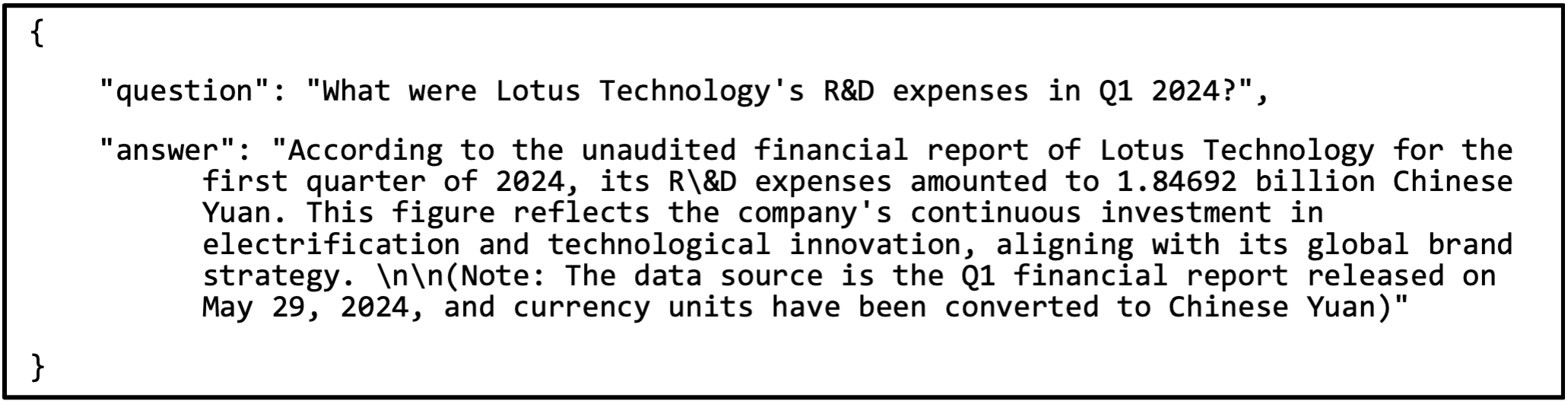}
    \caption{An example of question-answering using memory bank.}
    \label{fig:memory_bank_example}
    \vspace{-1em}
\end{figure}

The memory bank serves as a specialized, low-latency knowledge cache for fact-intensive queries. The illustrative example shown in Figure \ref{fig:memory_bank_example} shows that the key advantage of this mechanism is its ability to directly retrieve accurate, pre-computed answers for frequently asked questions, resulting in minimal end-to-end latency for high-frequency questions.

\section{Automatic Annotation} \label{app:annotation_prompt}
\subsection{Automatic annotation prompt template} 
\begin{figure}[h] 
    \centering 
    \includegraphics[width=1\linewidth]{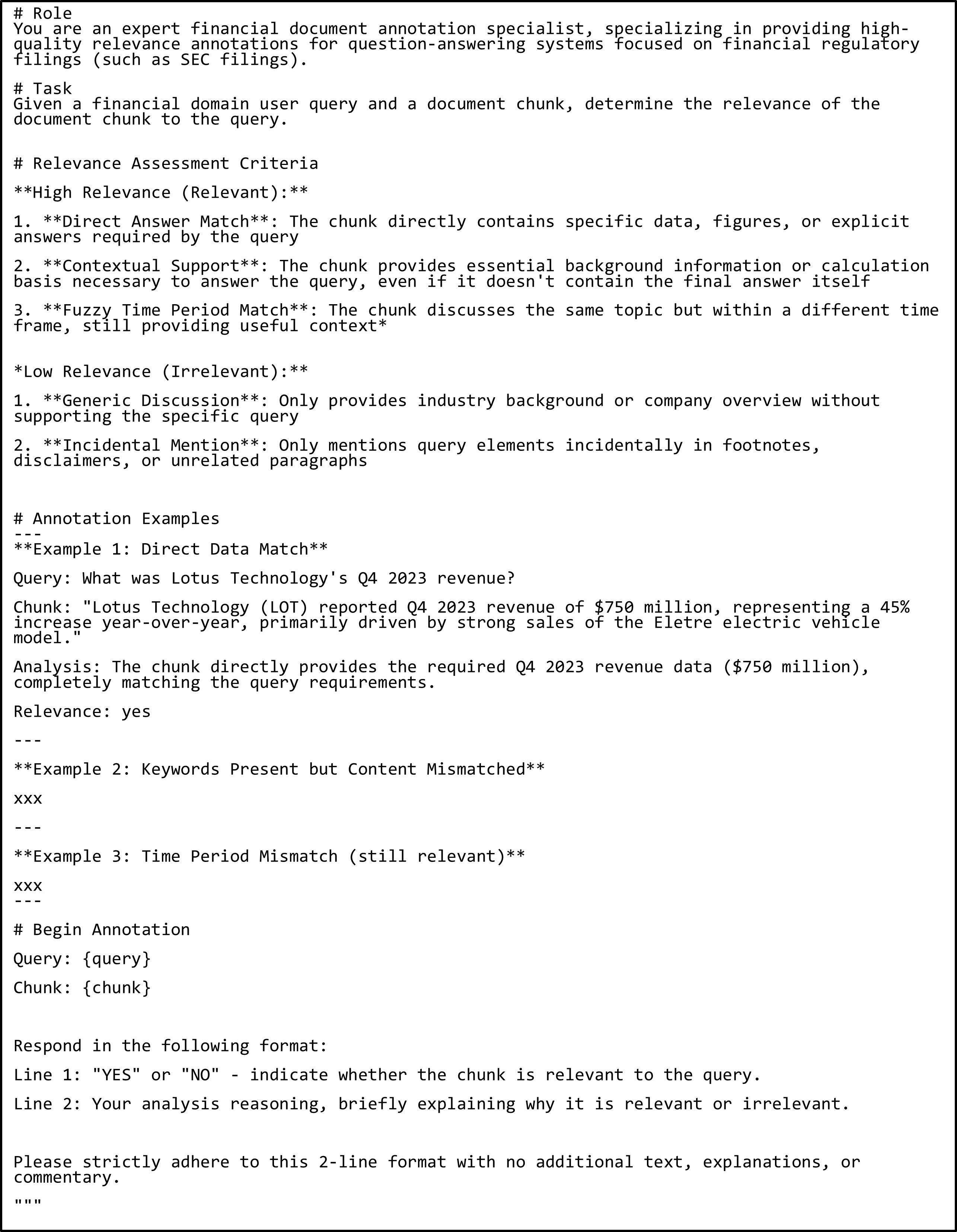}
    \caption{Automatic Annotation Prompt Template (Details partially omitted due to space constraints).}
    \label{fig:auto_annotate}
    \vspace{-1em}
\end{figure}

Figure \ref{fig:auto_annotate} presents the complete prompt template used for automatic labeling for retrieved document chunks. This template is designed to guide a Large Language Model (LLM), designated with the Role of an expert financial document annotation specialist, to perform the task of determining the relevance of a given document chunk to a user query. The prompt provides detailed Relevance Assessment Criteria, clearly defining both High Relevance (including Direct Match, Contextual Support, and Fuzzy Time Period Match) and Low Relevance criteria (Generic Discussion and Incidental Mention). Crucially, the prompt incorporates Few-Shot Annotation Examples to demonstrate the expected relevance judgment and accompanying analysis. Finally, it specifies a strict two-line output format to ensure structured and reliable data generation.

\subsection{Annotation accuracy}
To ensure the high quality and reliability of our in-house company datasets (Zeekr and Lotus), we adopted a rigorous annotation process based on our automatic annotation prompt template. In the first stage, we conducted a pilot annotation on a randomly selected subset of samples from both the Lotus and Zeekr datasets. The automatically generated annotations for these subsets were then subjected to manual review by domain experts. This manual quality check yielded high accuracy scores, confirming the initial effectiveness of the prompt. For challenging edge-case samples identified during this review, we further enhanced the prompt by integrating them as new few-shot examples. In the second stage, we leveraged this refined prompt template to annotate the entirety of the two datasets, ensuring consistent and high-quality relevance labels across the full data distribution. The quality assessment results are detailed in Table \ref{tab:annotation_accuracy}.
\vspace{-0.5em}
\begin{table}[htbp]
    \centering
    \caption{\textbf{Annotation Accuracy on Dataset Subsets.}}
    \label{tab:annotation_accuracy}
    \begin{tabular}{l c c}
        \toprule
        \textbf{Dataset} & \textbf{Subset Size (chunks)} & \textbf{Accuracy (\%)} \\
        \midrule
        Lotus & 1350 & 95.5 \\
        Zeekr & 1296 & 92.8 \\
        \bottomrule
    \end{tabular}
\vspace{-1em}
\end{table}

\section{Latency and Cost Analysis} \label{app:cost_analysis}
We conduct a detailed analysis of our system's operational efficiency, focusing on two critical aspects: \textbf{system latency} and \textbf{usage cost}. We first measure the average execution time for each process within our system. Crucially, we also quantify the average token usage for each call to the Chat Completion API, which serves as the primary factor in estimating the marginal deployment cost of the system. A deeper analysis is necessary for the asynchronous and variable components, particularly \textbf{Sub-query Answering} and \textbf{Tool Use}, since the system performance is dependent on the number of sub-queries executed and the number of available external tools. The system performance is detailed in Table \ref{tab:system_efficiency}.

\begin{table}[t]
    \centering
    \caption{System Latency and Deployment Cost. $t$ and $n$ ($n \ge 1$) denote the number of available tools and the number of sub-queries, respectively.}
    \label{tab:system_efficiency}
    \resizebox{\columnwidth}{!}{%
    \begin{tabular}{c l c c c c}
        \toprule
        \textbf{Step} & \textbf{Process} & \textbf{Time (s)} & \textbf{Model Used} & \textbf{Token Usage} & \textbf{Cost Estimation (per k queries)} \\
        \midrule
        1 & Query rewriting             & $\sim 2.4$ & DeepSeek-V3 API  & $\sim 600$  & $\sim \$0.17$ \\
        \midrule
        \multirow{2}{*}{2}
            & Retrieval \& re-ranking   & $\sim 1.8$ & Locally Deployed & N/A         & 24GB GPU \\
            & Tool Use (async, opt.)    & Varies     & DeepSeek-V3 API  & $\sim 250 \times n + 100 \times t \times n$ & $\sim \$0.07 \times n + \$0.02 \times t \times n$ \\
        \midrule
        3 & Sub-query answering (async) & $\sim 4.7$ & DeepSeek-V3 API  & $\sim 4500 \times n$ & $\sim \$1.3 \times n$ \\
        \midrule
        4 & Final answer merging        & $\sim 1.6$ & DeepSeek-V3 API  & $\sim 200 \times (n-1)$ & $\sim \$0.14 \times (n-1)$ \\
        \midrule
        \textbf{Total} & \textbf{--} & \textbf{$\sim 10.5$} & \textbf{--} & \textbf{$\sim 400 + 4.95\text{k} \times n + 100 \times n \times t$} & \textbf{$\sim \$0.11 + \$1.4 \times n + \$0.028 \times n \times t$} \\
        \bottomrule
    \end{tabular}
    }
\end{table}

\section{RAGAS} \label{app:ragas}
To comprehensively evaluate retrieval and generation performance, we employ the RAGAS framework \citep{es2025ragasautomatedevaluationretrieval}, which provides automated, reference-based metrics for RAG systems. 

\paragraph{Factual Correctness.} metric assesses the factual overlap between a generated response and a reference (ground-truth) answer. Internally, it uses an LLM as the evaluator to decompose both the response and the reference into sets of atomic claims. It then employs natural language inference to judge which response claims correspond to reference claims. A higher score indicates that the generated response preserves more ground-truth facts.

\paragraph{Answer Relevancy.}  
Answer Relevancy quantifies how pertinent or directly relevant the generated answer is to the user’s query. RAGAS implements it to penalize responses that are incomplete, extraneous, or tangential, awarding a higher score to answers that closely align in substance and focus with the original question.  

\paragraph{Context Recall.}  
Context Recall measures the extent to which the retrieved context covers all the claims needed to answer the question. In RAGAS, the reference answer is decomposed into claims, and the metric counts how many of those claims are supported by the retrieved context, using LLM-based attribution when necessary. Higher recall indicates fewer missing supports.

\paragraph{Context Precision.}  
This metric evaluates how well the retriever ranks relevant information above irrelevant passages. Specifically, it measures whether the ground-truth relevant items present in the contexts are ranked higher in the retrieve list, using a mean precision@k scheme over retrieved chunks. A high CP means that relevant chunks tend to appear at top ranks rather than buried.

\section{LightRAG and GraphRAG} \label{sub: graphrag-lightrag-config}
\subsection{GraphRAG Configuration}
GraphRAG is a graph-based retrieval system that uses knowledge graph representations to enhance contextual understanding. The configuration is detailed in Table \ref{tab:graphrag-config}. 

\begin{table}[t]
\centering
\begin{minipage}{0.9\linewidth}
\centering
\small
{\setlength{\fboxsep}{3pt}\setlength{\fboxrule}{0.6pt}%
\fbox{%
\begin{tabularx}{\linewidth}{@{}l >{\raggedright\arraybackslash}X@{}}
\multicolumn{2}{@{}l}{\textbf{LLM}}\\
Parameter & Value\\
\hline
Model & deepseek-v3\\
Message Caching&Enabled(hash-based KV)\\
\hline
\multicolumn{2}{@{}l}{\textbf{Embedding}}\\
Model & all-MiniLM-L6-v2\\
Dimension & 384\\
Max Token Size & 8192\\
Batch Size & 64\\
Max Async & 4\\
Normalize & True\\
\hline
\multicolumn{2}{@{}l}{\textbf{Retrieval / Query}}\\
Mode & local\\
Top cited chunks & 10 (EVAL\_TOP\_CHUNKS)\\
\hline
\multicolumn{2}{@{}l}{\textbf{Storage}}\\
Vector DB & NanoVectorDB\\
Graph Storage & NetworkX + GraphML\\
KV Storage & JSON\\
\end{tabularx}%
}}% end fbox group
\end{minipage}
\caption{Graphrag configuration}
\label{tab:graphrag-config}
\vspace{-1em}
\end{table}

\subsection{LightRAG Configuration}
LightRAG represents a more streamlined approach which prioritizes efficiency while maintaining retrieval quality. The detailed configuration is shown in Table \ref{tab:lightrag_config}.

\begin{table}[t]
\centering
\begin{minipage}{0.9\linewidth}
\centering
\small
{\setlength{\fboxsep}{3pt}\setlength{\fboxrule}{0.6pt}%
\fbox{%
\begin{tabularx}{\linewidth}{@{}l >{\raggedright\arraybackslash}X@{}}
\multicolumn{2}{@{}l}{\textbf{LLM}}\\
Parameter & Value\\
\hline
Model & deepseek-v3\\
Configuration & Custom LLM model func\\
\hline
\multicolumn{2}{@{}l}{\textbf{Embedding}}\\
Model & all-MiniLM-L6-v2\\
Dimension & 384\\
Max Token Size & 8192\\
Embedding batch num & 50\\
Max parallel insert & 2\\
\hline
\multicolumn{2}{@{}l}{\textbf{Query}}\\
Mode & global\\
Top\_k & 60\\
Response type & Multiple Paragraphs\\
Only need context & False\\
Max token for text unit & 4000\\
Max token for global context & 4000\\
Max token for local context & 4000\\
\end{tabularx}%
}}% end fbox group
\end{minipage}
\caption{LightRAG Configuration}
\label{tab:lightrag_config}
\vspace{-1em}
\end{table}

\end{document}